\documentclass[12pt]{article}
\def\simlt{\stackrel{<}{{}_\sim}}
\def\simgt{\stackrel{>}{{}_\sim}}
\def\bma#1{\mbox{\boldmath{$#1$}}}

\usepackage{amsmath}
\usepackage{amsfonts}
\usepackage{amssymb}
\usepackage{graphicx, rotating}
\usepackage{epstopdf}
\usepackage{epsfig}
\usepackage{latexsym}
\usepackage{graphicx}
\usepackage{color}
\usepackage{amsmath,bm,amssymb}
\usepackage{cite}
\usepackage{slashed}
\usepackage{hyperref}
\hypersetup{colorlinks, citecolor=bluscuro, linkcolor=black, urlcolor=bluscuro}
\definecolor{rossos}{cmyk}{0,1,1,0.55}
\definecolor{bluscuro}{rgb}{0.15, 0.2, .85}
\definecolor{bluchiaro}{cmyk}{1,.3,0.,0.1}
\definecolor{brown}{rgb}{0.6, 0.14, 0.14}

\newcommand{\beq}{\begin{equation}}
\newcommand{\eeq}{\end{equation}}
\newcommand{\beqa}{\begin{eqnarray}}
\newcommand{\eeqa}{\end{eqnarray}}
\newcommand{\bea}{\begin{eqnarray}}
\newcommand{\eea}{\end{eqnarray}}

\newcommand{\pp}{{\sigma}}

\newcommand{\eq}[1]{Eq.~(\ref{#1})}

\newcommand{\be}{\begin{equation}}
\newcommand{\ee}{\end{equation}}

\begin{document}

\begin{titlepage}
\vspace{.3in}

\vspace{1cm}
\begin{center}

\begin{center}
{\Large \bf 
Cosmological Higgs-Axion  Interplay\\
for   a  Naturally Small Electroweak Scale\vspace{0.1cm}\\
}
\end{center}
\vskip0.5cm

{\large J.R. Espinosa$^{a,b}$, C. Grojean$^{a,b,c}$,  
G. Panico$^{a}$,
A. Pomarol$^{d}$,\\
 O. Pujol\`as$^{a}$,~G. Servant$^{a,b,c,e}$}  \\[5mm]

{\it $^a$IFAE, Universitat Aut{\`o}noma de Barcelona, 08193 Bellaterra, Barcelona, Spain}\\[0mm]
{\it $^b$ICREA, Instituci\'o Catalana de Recerca i Estudis Avan\c{c}ats, Barcelona, Spain}\\[0mm]
{\it $^c$DESY, Notkestrasse 85, 22607 Hamburg, Germany}\\[0mm]
{\it $^d$Dept.~de~F\'isica, Universitat Aut{\`o}noma de Barcelona, 08193~Bellaterra,~Barcelona}\\[0mm]
{\it $^e$II. Institute of Theoretical Physics, University of Hamburg, D-22761 Hamburg, Germany}\\[0mm]
\end{center}
\bigskip

\vspace{.4cm}

\begin{abstract}
Recently, a new mechanism to generate a naturally  small electroweak scale has been proposed. It exploits the coupling of the Higgs to an axion-like field and a long era in the early universe where the axion unchains a dynamical screening of the Higgs mass.  We present a new realization of this idea with the new feature that  leaves no signs of new physics up to a rather large scale, $10^9$\,GeV, except for two very light and weakly coupled axion-like states. One of the scalars can be a viable Dark Matter candidate. Such a cosmological Higgs-axion interplay could be tested with a number of experimental strategies.
\end{abstract}
\bigskip

\end{titlepage}

\newpage 

\newpage

\section{Introduction}
\label{sec:Intro}

Our understanding of Nature is based on the empirical evidence that natural phenomena taking place at different energy/distance scales do not influence each other. At present, these different phenomena are described by a succession of effective theories with different  degrees of freedom manifesting themselves as shorter and shorter distances are probed. The parameters of the low-energy effective theory are {\it natural}  if they do not require any special tuning of the parameters of the theory at higher energies.

 Wilson~\cite{Wilson:1973jj} and 't Hooft~\cite{'tHooft:1979bh} gave a quantitative meaning to this naturalness principle by demanding that all dimensionless parameters controlling the different effective theories should be of order unity unless they are associated to the breaking of a symmetry. Numerous examples of the naturalness principle to understand the necessity of new phenomena have been extensively discussed in the literature (see for instance~\cite{Dine:2015xga} and references therein). 

The Higgs boson mass and the value of the cosmological constant have been long  recognized as two notorious challengers of this naturalness principle, a situation that stimulated the creativity of physicists in finding extensions of the Standard Model at higher energies. In most of these efforts to explain the smallness of the Higgs mass,  such as supersymmetric and composite Higgs models,
 new physics is predicted to  be present at TeV energies. 
Recently, however, a radically new approach to the Higgs mass hierarchy problem has been proposed~\cite{Graham:2015cka}, in reminiscence of the relaxation mechanism  of~\cite{Abbott:1984qf}  proposed for   explaining dynamically 
 the smallness of the cosmological constant (see \cite{Dvali:2004tma,Dvali:2003br} for similar previous ideas).
In principle, in this new approach no new degrees of freedom around the TeV scale are needed anymore to screen the Higgs mass from large quantum corrections. This has of course profound implications for the physics agenda of the LHC and beyond.

Technically, the relaxation mechanism of \cite{Graham:2015cka}    is based  on the cosmological interplay between the Higgs field $h$ and an axion-like field $\phi$,  arising from   the following three terms of the scalar effective potential:
\be
V(\phi,h)=\Lambda^3 g\phi- \frac{1}{2}\Lambda^2 \left(1-\frac{g\phi}{\Lambda}\right)h^2+\epsilon \Lambda_c^4 \left(\frac{h}{\Lambda_c}\right)^{n} \cos(\phi/f)+\cdots\, ,
\label{potential}
\ee
where  $\Lambda$  is   the UV cut-off   scale of the model,  while 
$\Lambda_c\lesssim \Lambda$ is the scale at which the periodic $\cos(\phi/f)$-term originates and
$n$ is a positive integer.
The first term  is needed to force  $\phi$ to roll-down in time, while  the  second  one corresponds to a   Higgs 
mass-squared term   with a (positive) dependence on $\phi$ such that
 different values of $\phi$   scan  the Higgs mass over a large  range, including the weak scale.
Finally, the third  term plays the role of a potential barrier for $\phi$,  dependent on  $h$, necessary to   stop the rolling of $\phi$ once  electroweak symmetry breaking (EWSB) occurs. 
For  this  mechanism to work, it is  also crucial to have a friction force, coming for example from Hubble friction during inflation, in order to make the rolling of $\phi$ very slow  to   access the right minimum during the cosmological evolution. 
We will discuss the possible ultraviolet (UV) origin of \eq{potential} later on.

At the classical level, the  proposed  mechanism can be understood  in the following way.
Assuming   that $\phi$ starts, at the beginning of the inflationary epoch, at  a  very large value
$\phi\simgt\Lambda/g$,  it will slow-roll   until it takes the critical value $\phi_c = \Lambda/g$, at which   the Higgs mass-squared becomes zero. From this  time on, as $\phi$ continues slowly rolling down,
the Higgs  mass becomes negative, and it is energetically favored to  turn on the  Higgs field.
This raises the  third term of \eq{potential} up to the point at which   $\phi$ stops  rolling.  For $g\ll 1$, this occurs for  a Higgs    value  $v$   given by
\be
g\Lambda^3\simeq \frac{\Lambda_c^{4-n} v^{n}}{f} \epsilon\, .
\label{hierarchy}
\ee
This equation arises from demanding that the  steepness of    the  linear $\phi$-term of the potential, first  term of \eq{potential}, equals the steepness of  the  Higgs barrier, third term of \eq{potential}.
From \eq{hierarchy} we see that we can have
 $v\ll \Lambda$  by taking $g$  small enough, which is technically natural as $g$   defines the spurion that breaks  the symmetry  
$\phi\to \phi+2\pi f$. 
At the quantum level, the described cosmological evolution is not much affected  provided
certain conditions,  specified in \cite{Graham:2015cka} and discussed later,
 are fulfilled.
Therefore, this mechanism  potentially offers  a new solution to the hierarchy problem.
We will refer to this  as   the cosmological Higgs-axion interplay (CHAIN) mechanism.

For $n=1$, the third term  of  \eq{potential} is linear in $h$, implying that
$\epsilon \Lambda_c^3$   must arise from a source of EWSB   other than the Higgs.
This   can be the QCD quark-condensate $\langle q\bar q\rangle\sim \Lambda^3_{QCD}$, 
as proposed in \cite{Graham:2015cka}. In this case  $\Lambda_c\sim\Lambda_{QCD}$
and $\epsilon\sim y_u$, where $y_u$ is the up-quark Yukawa.
This model, however, predicts too large a value for the QCD $\theta$-angle, in conflict with neutron electric dipole moment constraints.
A possible way to fix this problem was explained in \cite{Graham:2015cka}, but it requires a low cut-off scale, $\Lambda\simlt 30-1000$ TeV.
Alternatively, one could consider models in which the condensate comes from a new strongly-coupled sector,
a la Technicolor, with $\Lambda_c\sim$ TeV,
or  advocate the presence of an additional elementary Higgs doublet. 
This latter case however requires some extra symmetries  to keep the second Higgs doublet light, 
as we discuss in Appendix~\ref{app:TwoHiggsDoublet}.
These models predict extra physics carrying electroweak charges around the TeV scale that can be found in present or near-future experiments.

For $n=2$ on the contrary, the $h^2 \cos(\phi/f)$ term in $V(\phi,h)$  can arise from the  electroweak-invariant term $|H|^2\cos(\phi/f)$,  
where $H$ is the Higgs doublet,
and  therefore no extra source of EWSB  is needed beyond the SM Higgs. As a result,
these models can, in principle,   allow for a larger new-physics scale beyond the SM (BSM).
Nevertheless,   at the quantum level,  
extra terms can be now induced beyond those shown in \eq{potential}.
Indeed,  just    by closing   $H$ in a loop, we  expect,  at   $O(\epsilon)$, the  terms
\be
\epsilon\Lambda_c^4\cos(\phi/f)\ , \ \ 
\epsilon \Lambda_c^3\, g\phi\cos(\phi/f)\, ,
\label{extra}
\ee
to be  generated. These   terms   give a potential to $\phi$ that,  unless $\Lambda_c\lesssim v$,
make it stop  slow-rolling much before  the Higgs   turns on.
Therefore, if we want the CHAIN  mechanism to work,  we must have again new physics  not far away from the weak scale
and therefore potentially visible in forthcoming experiments.
It is important to notice that this  new physics is not responsible for keeping the Higgs light,
but for generating the periodic term of \eq{potential}.
In the particular model of this type  discussed in \cite{Graham:2015cka},  extra fermions were predicted  at around the weak scale.
An important drawback of this type of  models is that they must address a ``coincidence problem":
 they must provide a  BSM scale $\Lambda_c$  that must lie  around the weak scale with no  a priori reason,  as   the weak scale is determined by \eq{hierarchy}.

The aim of our work is to offer an existence proof that it is indeed possible to devise a model that dynamically generates a large mass gap between the Higgs mass and the new physics threshold. 
The proposed model will not have a ``coincidence problem" as the only new physics scale
will be associated with $\Lambda\sim \Lambda_c \gg v$.
For this to work, we  need to make the terms of \eq{extra} smaller than the term  $\epsilon\Lambda^2 |H|^2\cos(\phi/f)$. 
For this purpose, we will   introduce another slow-rolling field, $\pp$,  coupled to 
$\cos(\phi/f)$. During its  cosmological evolution,   $\pp$  will take a  value such that  $\pp\cos(\phi/f)$ will  cancel  the   
 terms of \eq{extra}.  When this occurs,  $\phi$ will be free to move, tracking $\pp$ downhill. 
Only when the $h$-dependent term turns on, $\phi$ will  stop tracking $\pp$ and reach the minimum
fixed by \eq{hierarchy}.
We will be able to push the cut-off scale up to  $\Lambda\sim 10^9$\,GeV, providing the first example of a natural theory
with such a large BSM scale. 
The only  new   states,  $\phi$ and $\sigma$, will have masses below the weak scale, but they will be 
very weakly coupled to the SM, making them very difficult to detect at present and future experiments.
Interestingly, as we will see, they could  provide the source of dark matter needed in the universe.

\section{Double scanner mechanism\label{mechanism}}
\label{sec:model} 

The key new ingredient   of  our  proposal,
with respect to \cite{Graham:2015cka},  
 is a  second scanning field, that we call $\pp$. The full potential, up to terms of order $\epsilon$, $g_\pp$ and $g$, 
 is given by
 \be
	\label{eq:V}
V(\phi,\pp,H) = \Lambda^4 \left( \frac{g \phi}{\Lambda} +   \frac{g_{\pp} \pp}{\Lambda}\right) -\Lambda^2\left(\alpha -   \frac{g \phi}{\Lambda}\right) |H|^2 +\lambda |H|^4 +A(\phi,\pp,H) \cos{\left(\phi/  f\right)}\ ,
\ee
where 
\be
A(\phi,\pp,H) \equiv \epsilon \Lambda^4 \left(\beta + c_\phi \frac{g \phi}{\Lambda}  -c_\pp  \frac{g_\pp\, \pp}{\Lambda}  +  \frac{|H|^2}{\Lambda^2} \right)\ ,
\label{envelope}
\ee
with $0<g,g_\pp,\epsilon\ll 1$, while
$\alpha, \beta $ and $c_{\phi},c_\pp$ are $O(1)$ positive coefficients. 
 We assume that all terms of \eq{eq:V} are generated at the cut-off scale $\Lambda$.
For simplicity and clarity,  we are only  considering  linear terms in $g \phi /\Lambda$, but
 we could  have taken a generic  function of $g \phi /\Lambda$    with the only requirement that it is 
 monotonically decreasing or increasing in a wide region of order $\Lambda/g$ (and similarly for $\pp$ with $g\rightarrow g_\pp$).

From Eqs.~(\ref{eq:V}) and~(\ref{envelope}), we see that $\phi$ scans the Higgs mass, while  $\pp$ scans $A(\phi,\pp,H)$, the overall amplitude --the {\em envelope}-- of the oscillating  term.
This dependence of $A(\phi,\pp,H)$  on $\pp$ is  crucial  for our CHAIN mechanism to work, while the other terms in \eq{envelope} are added since, as we said, they are anyway generated  at the  quantum  level (by loops of~$H$).
  The potential in~\eq{eq:V} is   stable under quantum corrections in the small-coupling limit  ($g,g_\pp,\epsilon\ll 1$)
we consider. A possible UV origin of the periodic term in~\eq{eq:V} is given in Appendix A.

\begin{figure}[t]
$$ \hspace{-.5cm}\includegraphics[width=0.55\textwidth]{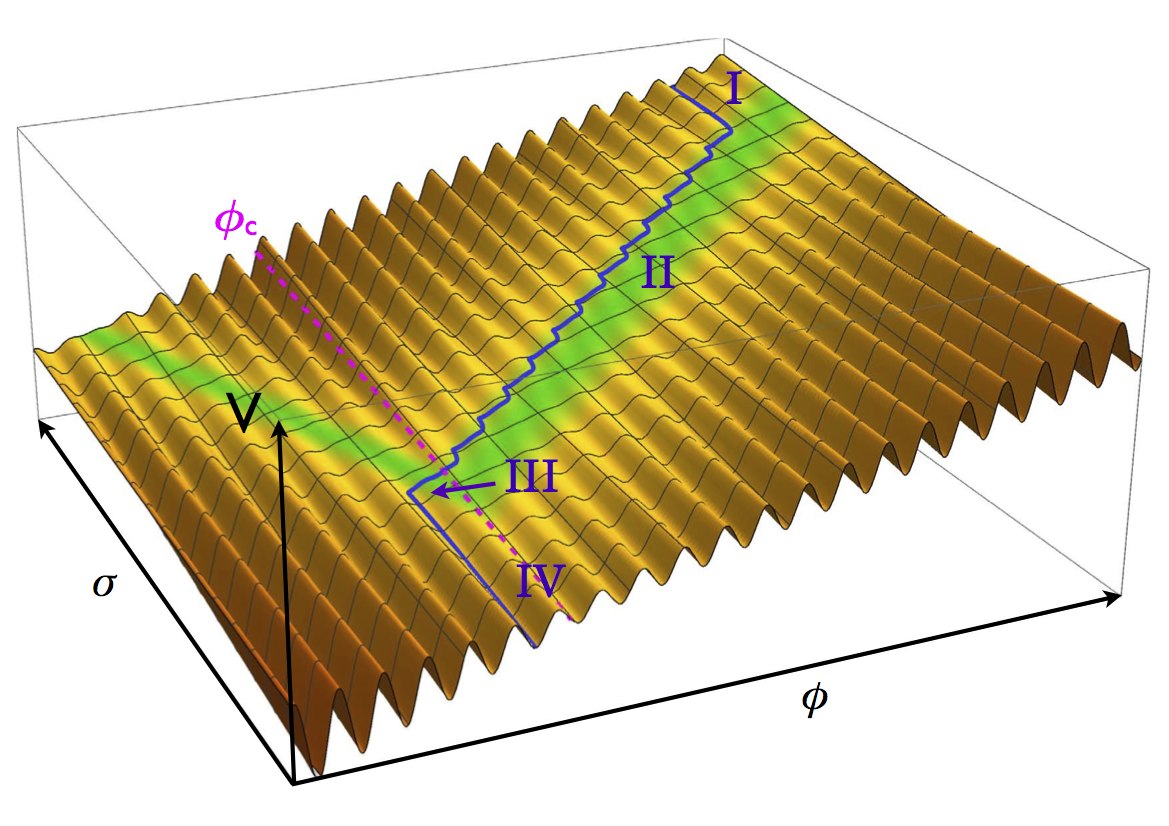} \hspace{1.cm}
\includegraphics[width=0.4\textwidth]{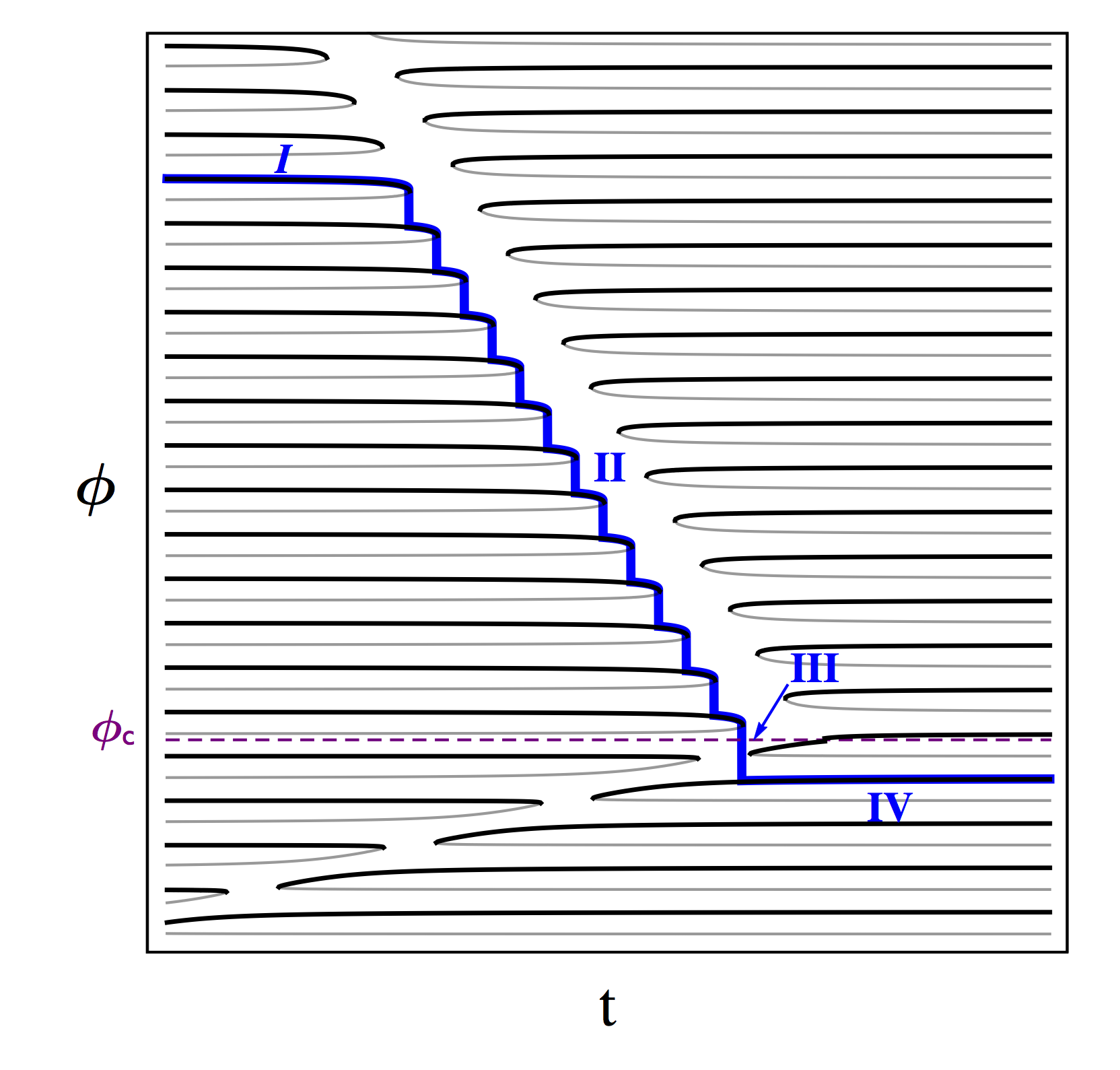}  $$  
\begin{center}
\caption{\emph{ {\bf Left}: Scalar potential in the  $\{\phi,\pp\}$ plane.  The band without barriers is in green while the barriers getting high(er) are dark(er) brown. The blue line shows a possible slow-roll cosmological trajectory of the fields during inflation. The dashed purple line is the critical line for EWSB.
{\bf Right}: Classical time evolution of $\phi$ (blue curve) in the potential on the left. The black lines show the extremal points of the potential, with closely spaced minima (bold) and maxima (thin) alternating.  
(Arbitrary units and scales in both plots.) \label{fig:V2Dphit}}}
\end{center}
\end{figure}

\begin{figure}[!t]
$$\includegraphics[width=0.5\textwidth]{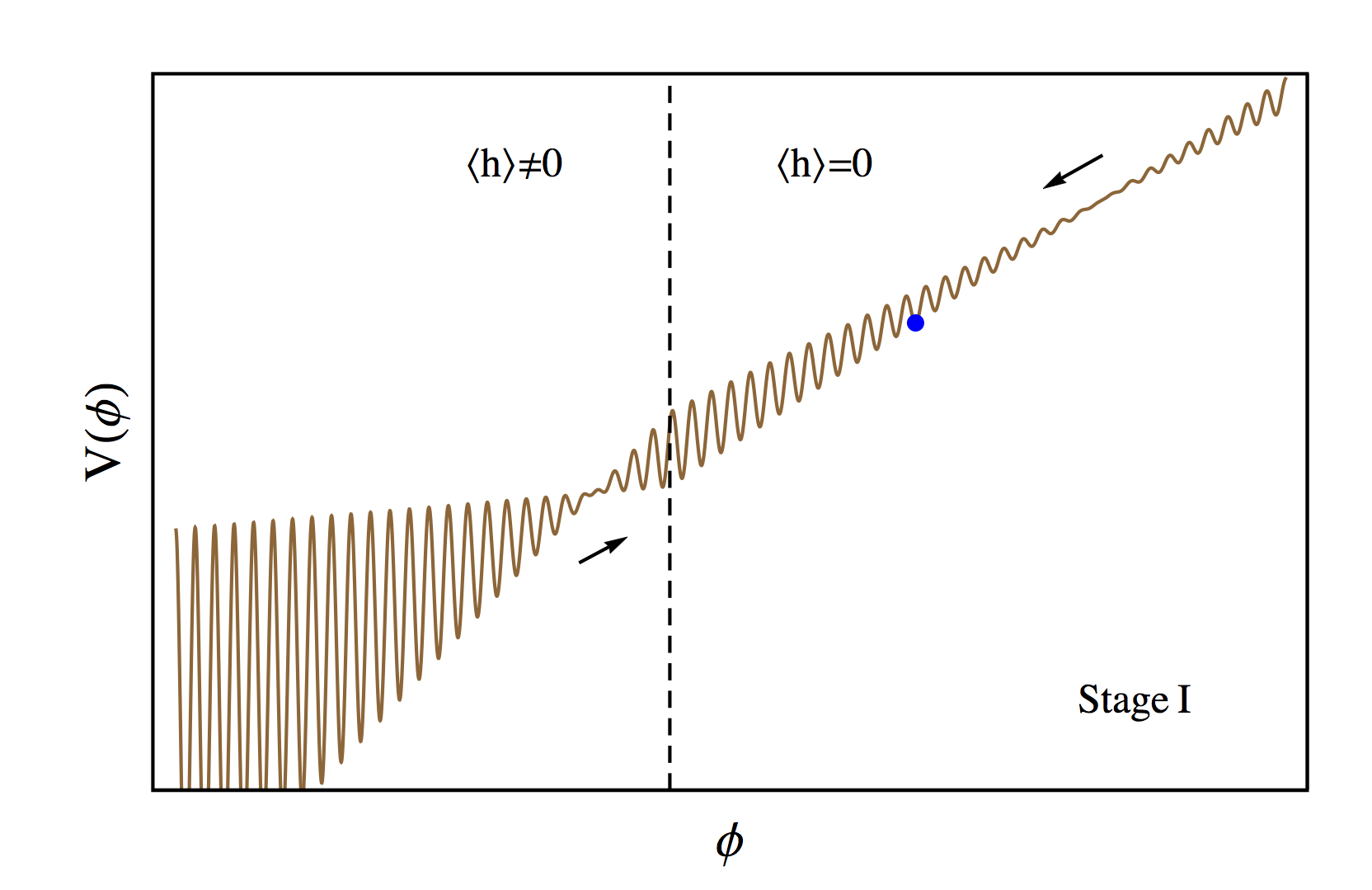} \hspace{.2cm}
\includegraphics[width=0.5\textwidth]{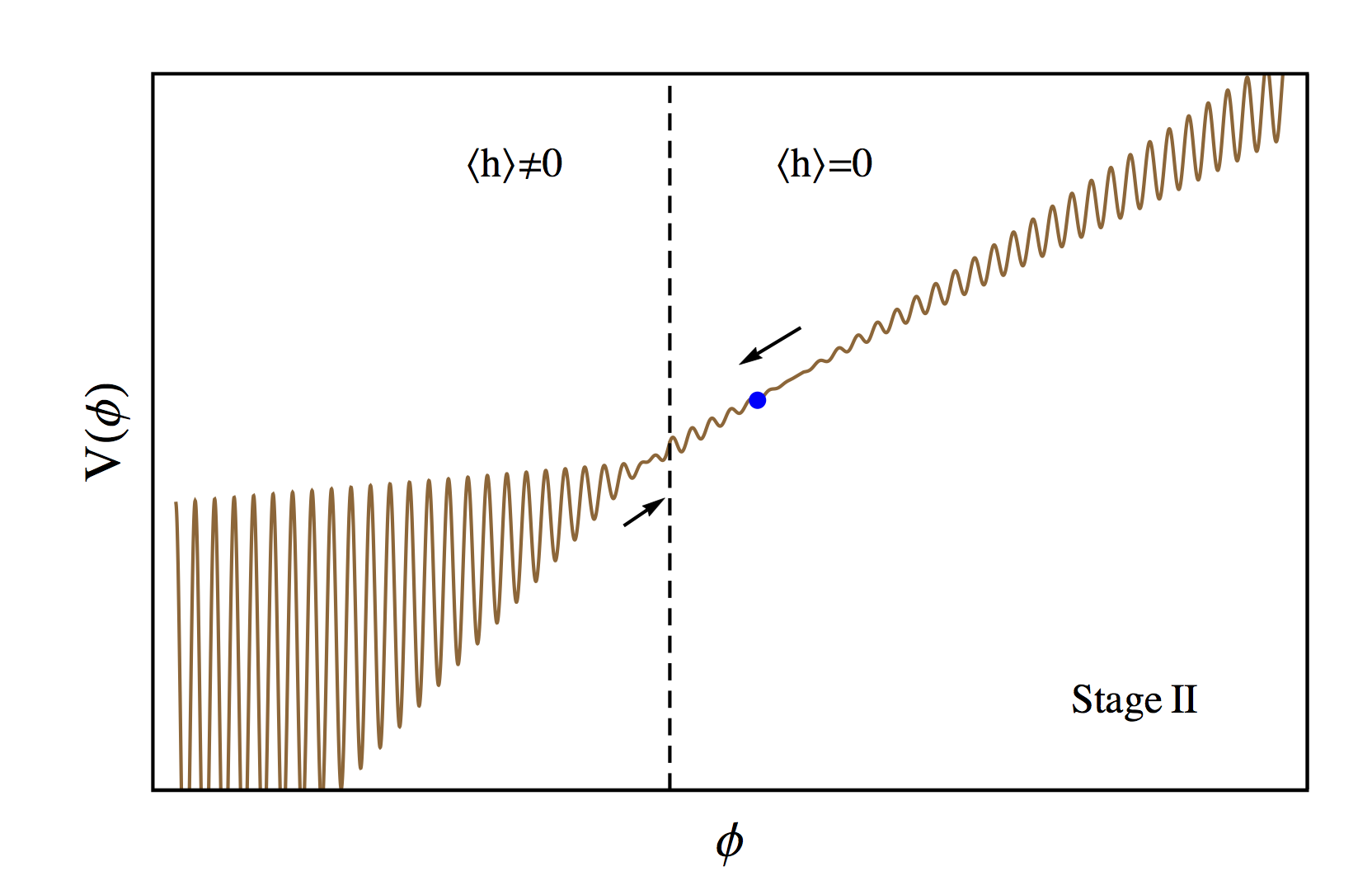} $$ 
$$\includegraphics[width=0.5\textwidth]{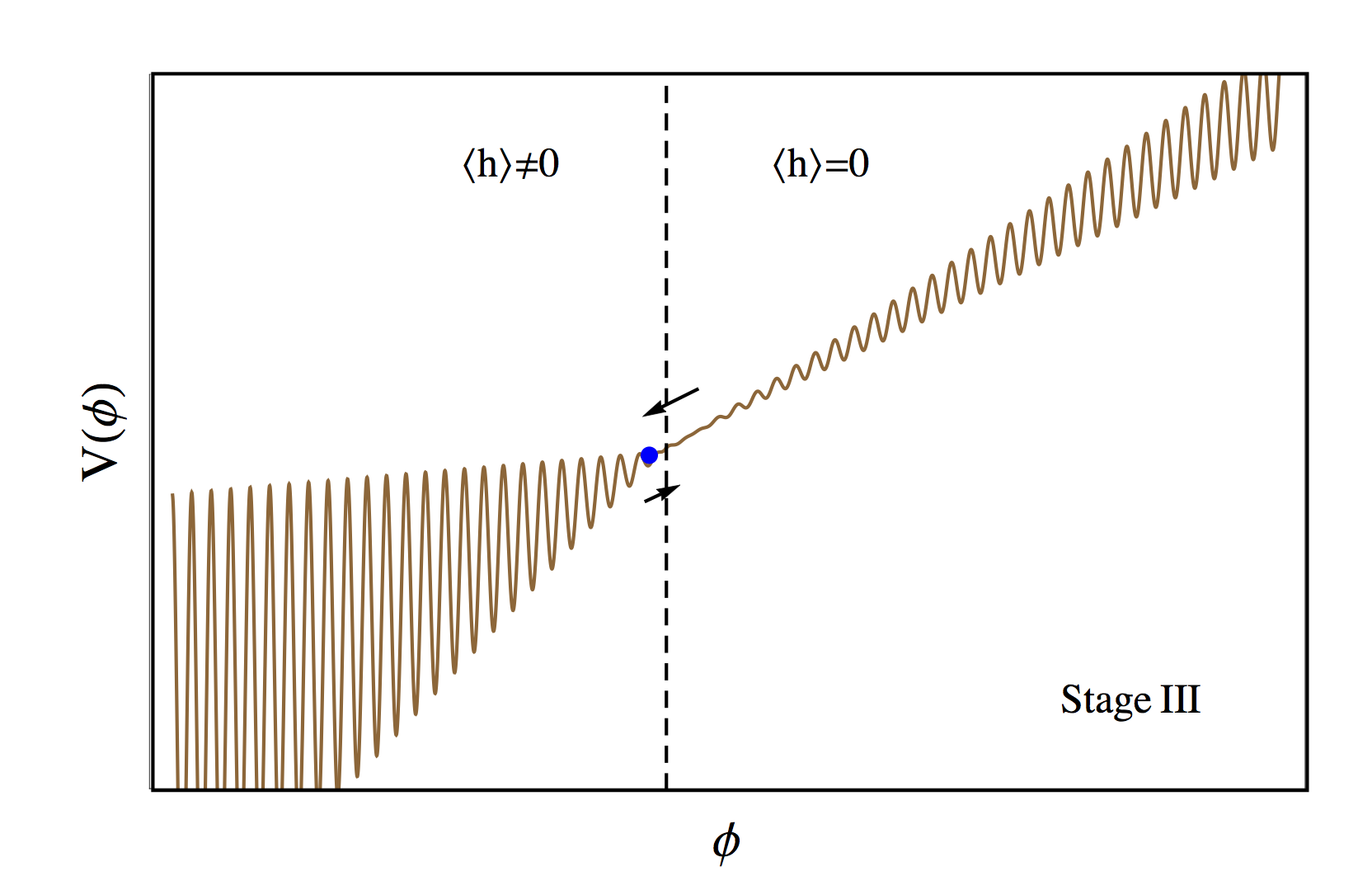} 
\hspace{.2cm}
\includegraphics[width=0.5\textwidth]{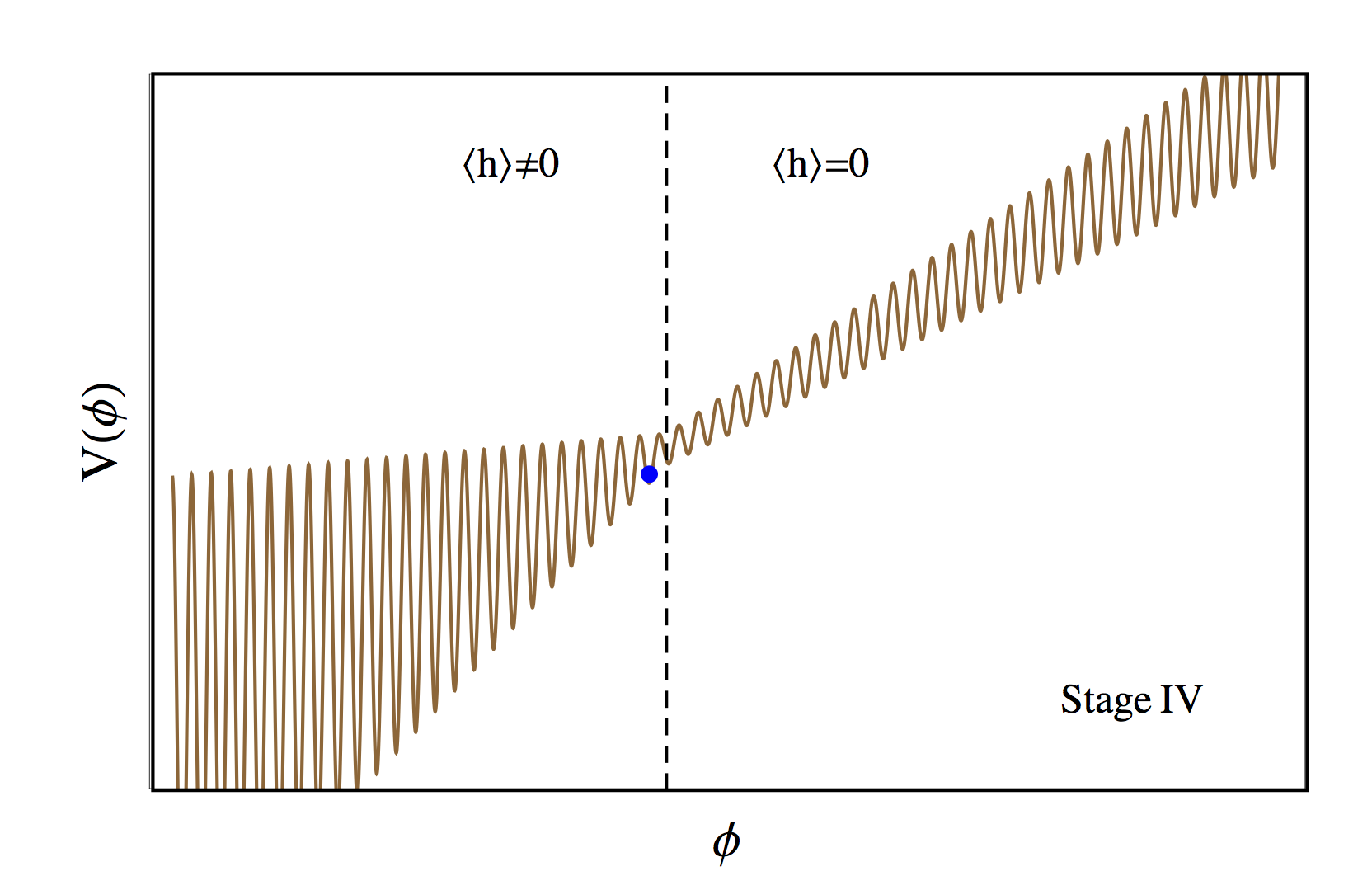} $$ 
\begin{center}
\caption{\emph{Sketch of the four stages in the evolution of $\phi$, marked by the blue dot, in the time-dependent effective potential for $\phi$ obtained after integrating out $\sigma$ and $H$ but corresponding to the same potential as in Fig.~\ref{fig:V2Dphit}. (Arbitrary units and scales.) }\label{fig:stages}}
\end{center}
\end{figure}

We will study the time evolution of $\phi$, $\pp$ and $H$ during the inflationary epoch.
Inflation is needed, as in \cite{Graham:2015cka}, to  provide  the friction that  makes the fields  slow-roll  and reach the desired minimum. The time evolution of $\pp$ is quite trivial, as for $\epsilon\ll 1$, it  simply slides down: 
\be
\pp (t)= \pp_0 - g_\pp \Lambda^3 t / (3 H_I)
\ .
\ee
In the cosmological evolution of $\phi$  we can   distinguish four stages, depicted in Fig.~\ref{fig:V2Dphit}, that we qualitatively describe next:
\begin{enumerate}
\item[I)] At the start of inflation we assume   $\phi\gtrsim    \Lambda/g$   and  $\pp\gtrsim  \Lambda/g_\pp$  such that the
Higgs mass-squared and the amplitude $A$ are positive.  
The field $\phi$ is stuck  in some  deep minimum coming from the $A\cos(\phi/f)$ term 
of \eq{eq:V}, while the Higgs field value is zero.

\item[II)]  As   $\pp$ evolves down,    the amplitude $A$ decreases until  the point at which for $\phi$ 
the steepness  of $A\cos(\phi/f)$ is smaller than  the slope coming from  the linear term  of \eq{eq:V},  and $\phi$ can start to    move down.
The region  in field-space at which this occurs is shown by a ``green-band"  in 
Fig.~\ref{fig:V2Dphit}. In this region, the bumps from $A\cos(\phi/f)$
are very small and,  for $g_\pp\lesssim g$,    $\phi$ goes down tracking $\pp$: $\phi(t)\simeq   \textrm {const.}+c_\pp g_\pp\pp(t)/(c_\phi g)$,
which is the solution of $A\approx 0$ (this solution neglects effects of size $\Delta \phi \sim f$ which correspond to the stepwise behavior visible in Fig.~\ref{fig:V2Dphit}).  

\item[III)] When   $\phi$ crosses  the critical value $\phi_c \equiv \alpha\Lambda/g$
the Higgs mass-squared term becomes negative, turning on   $H$.  This gives,
according to   \eq{envelope},  a positive  contribution to   the amplitude  $A$, that, 
for certain values of the parameters of \eq{eq:V} to be specified later, 
makes  the direction of the green-band   change  as shown in   Fig.~\ref{fig:V2Dphit}.
The field $\phi$  cannot continue any longer its evolution along  the tracking trajectory that would bend in the opposite direction in $\sigma$ and would not be compatible with the cosmological evolution of $\pp$ that imposes $d\pp /dt<0$. So $\phi$ crosses the green-band.  
 
\item[IV)] Finally,  $\phi$ reaches the other side of the   green-band (where $A$ becomes large again) 
and gets stuck in a minimum from $A\cos(\phi/f)$ as in the model of \cite{Graham:2015cka}.
 The field $\pp$ continues going down,  making  $A$ grow    until $\pp$ finds its own minimum.
\end{enumerate}

For the same potential shown in Fig.~\ref{fig:V2Dphit}, we show in Fig.~\ref{fig:stages} four snapshots of how $\phi$ evolves in the time-dependent potential $V(\phi)\equiv V(\phi, \pp(t),h(\phi))$, obtained after integrating out $\sigma$ and $h$.
We are  choosing four representative time values corresponding to the four stages I-IV. 
At the stage I and II, this potential has two  $A\approx 0$ regions  moving towards each other (as  indicated by the arrows),
that merge  at stage III and disappear at stage IV.

To understand under which conditions the potential of \eq{eq:V}  has the shape shown in Fig.~\ref{fig:V2Dphit},
we start by     finding  the   $\phi$ values, called  $\phi_*$, for which  the steepness from $A\cos(\phi/f)$ is smaller than the steepness from the  $\Lambda^3 g\phi$ term of \eq{eq:V}. These are determined by 
\be
\frac{1}{f}A(\phi_*,\sigma,h(\phi_*))\lesssim g\Lambda^3\, ,
\label{phistart}
\ee
where we are working in the limit $g,g_\pp\ll 1$. In this range of $\phi$ values, $\partial V/\partial \phi=0$ has no solutions, leading
to a ``sliding" region, the minima-free green-band of Fig.~\ref{fig:V2Dphit}.

The value of the neutral Higgs component, $\langle H \rangle = ( 0,  h/\sqrt{2})$, depends on $\phi$ according to\footnote{To simplify the discussion,  we are ignoring a term $-(\epsilon \Lambda^2/\lambda) \cos(\phi/f) $ in the RHS 
of \eq{h2phi}, that causes a transitory period in which the Higgs mass term switches sign repeatedly before stabilizing as negative.\label{foot:BackreactionHiggsMass}}
\be
	\label{h2phi}
h^2(\phi) \simeq 
\displaystyle \frac{\Lambda^2}{\lambda}\left(\alpha  - \frac{g \phi}{ \Lambda}\right)\  ,\   \rm{for}\; \phi<\phi_c\,  ,
\ee 
and zero otherwise. 
Solving  \eq{phistart}, we obtain   the interval(s)
\be
\phi_*\in \left\{
\begin{array}{ll}
\phi_c+ \displaystyle\frac{c_\pp\, g_\pp}{c_\phi\, g}\left(\pp-\pp_c\right)\pm \frac{f}{c_\phi\, \epsilon }\, 
, &(\rm{for}\; \phi_*>\phi_c
)\\ &\\
\phi_c+\displaystyle\frac{c_\pp\, g_\pp}{c'_\phi\, g}\left(\pp-\pp_c\right)\pm \frac{f}{c'_\phi\, \epsilon }\,   
, & (\rm{for}\; \phi_*<\phi_c )\, ,
\end{array}
\right.
\label{interval}
\ee
with $c'_\phi = c_\phi-1/(2\lambda)$.
 By continuity, both solutions merge at $\phi_*=\phi_c$ at a particular value $\sigma_c= (g c_\phi \phi_c+\beta\Lambda)/(c_\pp g_\pp)$.\footnote{In fact, the kink in the sliding-band is located around the point $\{\phi_c,\sigma_c\}$. In terms of these coordinates one has $A=\epsilon\Lambda^3[c_\phi g(\phi-\phi_c)-c_\pp g_\pp(\pp-\pp_c)+|H|^2/\Lambda]$.}


In order for $\phi(t)$ to track down $\pp(t)$, or what is equivalent, for $\phi(t)$ to stay  in the  $\phi_*$ region of \eq{interval} until reaching  $\phi_c$, the gradient of the 
dynamical trajectory  in  the $\{\phi,\pp\}$ plane\footnote{Inside the $\phi_*$ interval, we have $A\approx0$ and  $\phi$'s slow-roll is driven by the linear term of \eq{eq:V}.} inside the $\phi_*$  interval,
\be
\frac{d\phi(t)/dt}{d\pp(t)/dt}=\frac{g}{g_\pp}\, , 
\ee
should be larger than  the gradient ${d\phi_*}/{d\pp}$ of the green-band itself.
Otherwise 
$\phi(t)$ crosses the green-band too early and gets stuck at some minimum before getting to $\phi_c$.
From \eq{interval}, this condition, ${d\phi(t)}/{d\pp(t)}>{d\phi_*}/{d\pp}$,
 leads to the requirement
\beq\label{tracking}
c_\phi g^2  > c_\pp g^2_\pp\ .
\eeq
On the contrary, once  $\phi\leq \phi_c$  we must demand $\phi$ to exit the green-band (getting trapped in some vacuum precisely as needed to explain the smallness of the electroweak scale)
which implies $d\phi(t)/d\pp(t)<d\phi_*/d\pp$ and leads to
\be
 c'_\phi g^2 < c_\pp g^2_\pp\ .
\label{con2}
\ee
This condition is easily satisfied for  $c'_\phi <0$, which is equivalent  to $ 2\lambda c_\phi<1$,  and corresponds to the situation depicted in 
Fig.~\ref{fig:V2Dphit} in which the green-band flips direction at $\phi=\phi_c$.
For $ 2\lambda c_\phi>1$, the green-band does not switch direction at $\phi_c$ but its slope changes. This occurs however  for a small range of $c_\phi$ where \eq{con2} is also fulfilled:
\be
\frac{1}{2\lambda} < c_\phi < \frac{1}{2\lambda}+\frac{c_\pp g_\pp^2}{c_\phi g^2}\, .
\ee 
This  region of the parameter space  can  then also provide  an explanation to  the smallness of the electroweak scale.

Let us emphasize that the CHAIN mechanism described above works independently of  $\phi_i$, the initial condition for $\phi$, provided only that $\phi_i$ at the initial time $t_i$ lies in the region $\phi_c < \phi_i < \phi_*(t_i)$, which is a natural and sizable  range of the available field space.
A  $g_\pp \pp  |H|^2$ term in the potential could spoil the CHAIN mechanism,
as the late evolution of  $\pp$, after $\phi$ is settled in its minimum,   would  change the value of 
the Higgs mass. Therefore  we find that this term, if present, must be further suppressed by an extra factor  $\epsilon$.
This does not pose any problem to our  model as 
a radiatively generated $\pp  |H|^2$ term   arising from \eq{eq:V}  by loops of $\phi$, 
is  indeed   very small,  of $O(\epsilon^2 g_\pp)$.

\section{Consistency requirements for a  small  weak scale}
\label{sec:Consistency}

The potential in \eq{eq:V} involves two scales, $\Lambda$ and $f$, and three small couplings, $g, g_\pp$ and $\epsilon$, apart from  a few dimensionless parameters of order unity (including the Higgs quartic coupling, $\lambda$). The cosmological evolution of our model can be broadly described by these  parameters  and two external quantities fixed by the inflaton sector: $H_I$, the value of the Hubble parameter during inflation, and $N_e$, the number of $e$-folds. These various parameters have to satisfy several constraints in order to provide a natural solution to the hierarchy problem, i.e., to dynamically ensure a  stable separation between the weak scale and the high-energy scale $\Lambda$  from which the potential \eq{eq:V} emerges as a low-energy description. These constraints  arise from demanding

\begin{enumerate}

\item {\it Dangerous quantum corrections to the potential are kept small.} At least at the quantum level,  additional terms
 of $O(\epsilon^2)$  can be generated in the potential of \eq{eq:V} that could  potentially spoil our CHAIN mechanism.
For instance,   terms like\footnote{See  Appendix A for the  possible origin of these terms in a particular UV completion.} $\epsilon^2\Lambda^4\cos^2(\phi/f)$ or $\epsilon^2 \Lambda^3 g\phi\cos^2(\phi/f)$, 
depend quadratically  on $\cos(\phi/f)$,
and therefore  their amplitudes cannot be cancelled  by $\pp$ simultaneously to 
$A\cos(\phi/f)$.
 These terms are dangerous since  they give a barrier  to $\phi$ at   values that can be above the critical $\phi_c$.
To make sure that they remain subdominant  to the Higgs barrier of \eq{eq:V}, we must demand
\be
\label{eq:quantumstability}
\epsilon \lesssim v^2/\Lambda^2\, .
\ee
This condition also ensures that the contribution to the Higgs mass coming from the $\epsilon \Lambda^2 |H|^2 \cos(\phi/f)$
term in the potential is at most of  electroweak size and does not spoil the tracking  behaviour (see
footnote~\ref{foot:BackreactionHiggsMass}).

\item {\it $\phi$ must be trapped by the Higgs barrier.} As in the model of  \cite{Graham:2015cka},
  the nonzero Higgs field must be the  only one responsible for stopping $\phi$ from sliding any longer.
  This is the requirement  in \eq{hierarchy} that, for our case $n=2$ and $\Lambda=\Lambda_c$, reads $g \Lambda^3  \simeq \epsilon \Lambda^2 v^2/f$.
This can be used to obtain the  electroweak scale as a prediction from the model in terms of microscopic parameters:
\be
\label{eq:endtracking}
v^2
 \simeq 
 \frac{g \Lambda f}{\epsilon }\, .
\ee
We will also use this relation later on to get rid of $\epsilon$ in terms of the other parameters of the model, so that the electroweak scale is reproduced correctly.

\begin{figure}[!ht]
$$\includegraphics[width=0.7\textwidth]{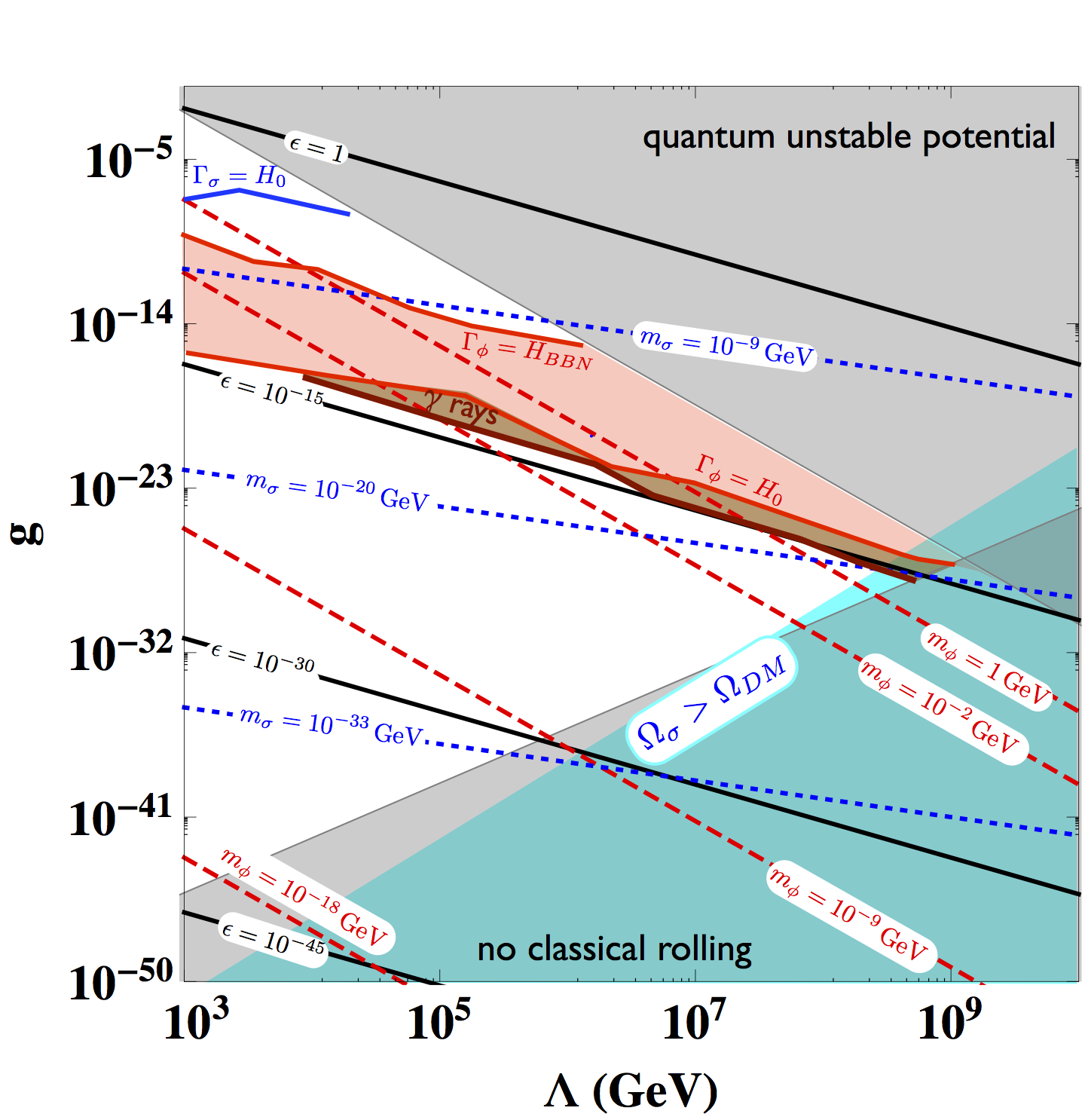}  $$  
\begin{center}
\caption{\emph{Parameter space for a successful solution of the hierarchy problem ensured by the cosmological evolution of the fields $\phi$ and $\pp$. We have taken $\Lambda=f$ and $g_\pp=0.1g$.
In the upper shaded gray region, the quantum corrections to the potential would not drive the system towards the weak scale while in the lower shaded gray region, the two fields would not follow their classical paths. Some contour lines of the expansion parameter $\epsilon$ (continuous black) and of the physical masses of $\pp$ (dotted blue) and $\phi$ (dashed red) are  reported. In the shaded red region, the field $\phi$ decays after BBN and has a lifetime shorter than the age of the universe.The brown shaded region is excluded by gamma-ray data. Finally, in the shadded cyan region on the bottom right corner, the energy density stored in the oscillations of the field $\pp$ exceeds the DM energy density.
}\label{fig:gversusLambda}}
 \end{center}
\end{figure}

\item {\it Inflation is independent of the $\phi$ and $\pp$ evolution.}  We assume for simplicity that   inflation is  driven by another field, the inflaton, that does not receive  any back-reaction from the evolution of  $\phi$ and $\pp$.
 This is possible under the condition that the typical energy density carried by $\phi$ and $\pp$ remains smaller than the inflation scale, i.e.,
\be
\label{eq:inflation}
\frac{\Lambda^2}{M_P}\lesssim H_I \, ,
\ee
where $M_P$ is the reduced Planck mass, $M_P\simeq 2.4\times 10^{18}$\,GeV.
\item {\it Classical rolling dominates over quantum jumping.}  We are assuming that  the cosmological evolution of  $\phi$ and $\pp$ is  dominated by  classical physics. It is therefore essential, for the consistency of our solution, that during the cosmological evolution of our system the quantum fluctuations of the fields, typically of  size $H_I$, remain smaller than the classical field displacements over one Hubble time, i.e., $\Delta \pp \sim H_I^{-1} d\pp /dt \sim H_I^{-2} dV /d\pp$ for the case of $\pp$. This condition for classical rolling  \cite{Graham:2015cka}
 simply reads
\be
\label{eq:classicalrolling}
H_I^3\lesssim  g_\pp \Lambda^3\, ,
\ee
 for $\pp$. Due to  \eq{tracking}, the classical-rolling condition for $\phi$ is  automatically guaranteed whenever \eq{eq:classicalrolling} is fulfilled (we assumed that $c_\phi\sim c_\pp\sim 1$). 


 \item {\it Inflation lasts long enough for complete scanning.} In order for the above   mechanism to work for  generic initial field configurations, it is essential that the range scanned by $\phi$ and $\pp$ during inflation  be of order or  larger than $\Lambda/g$ and $\Lambda/g_\pp$ respectively. This is ensured by requiring a long enough period of inflation, namely,
$N_e\Delta\pp\sim N_e H_I^{-2} {dV /d\pp}\gtrsim \Lambda/g_\pp$,  which leads to
  \be
\label{eq:longinflation}
N_e \gtrsim\frac{H^2_I}{g_\pp^2 \Lambda^2}\, .
\ee
The typical duration of the different stages of the cosmological evolution of
$\phi$ is of the same order, with the exception of stage III, which is much shorter, of order $N_e^{III}\sim (gf/\epsilon \Lambda) N_e$. 
  \end{enumerate}
Combining these various consistency conditions,  together with \eq{tracking},
we obtain that the couplings $g_\pp$ and  $g$ are bounded to the interval:
\be
\frac{\Lambda^3}{M_P^3} \lesssim g_\pp\lesssim g \lesssim \frac{v^4}{f \Lambda^3}.
\label{bound}
\ee
Since $f$ cannot be much smaller than $\Lambda$, as this latter is the scale at which the $\cos(\phi/f)$ term is generated,
 we obtain from \eq{bound} an upper bound on the cut-off of our model:
\be
 \Lambda \lesssim (v^4 M_P^3)^{1/7} \simeq 2\times 10^9\,\textrm{GeV}\, .
 \ee
The bound in \eq{bound}
 defines the region of the parameter space of the model consistent with a natural solution to $v\ll \Lambda$.
This is shown in  Fig.~\ref{fig:gversusLambda}, where, for concreteness,
 we have taken  $f= \Lambda$ and  $g_\pp = 0.1  g$.

\section{Quantum spreading}
\label{sec:spread}

The discussion of the cosmological evolution in Section~\ref{mechanism} was fully classical. As we saw in more detail in the previous Section, the model parameters can be chosen to ensure that the slow-roll of the fields $\phi$ and $\pp$ is indeed dominated by classical evolution, with quantum fluctuations (driven by $H_I$) playing a subdominant role. As discussed in \cite{Graham:2015cka}, however, one should pay attention to the subleading quantum spreading to ensure that it does not spoil the solution to the hierarchy problem.

This issue can be attacked using a Langevin equation (that adds
to the usual classical evolution of the field a noise term that captures the random fluctuations controlled by $H_I$) or using the Fokker-Planck equation\footnote{For more details on the use of these equations to describe fluctuating light fields (like the Higgs) during inflation, see {\it e.g.}~\cite{HdI}.}  
\be
\frac{\partial P}{\partial N_e} = \frac{\partial}{\partial\phi}\left(\frac{P}{3H_I^2}\frac{\partial V}{\partial\phi}\right)+\frac{\partial^2}{\partial \phi^2} \left(\frac{H_I^2}{8\pi^2}P\right) \ ,
\label{FP}
\ee
for $P(\phi,N_e)$ the probability (density) of finding the field in the interval $(\phi,\phi+d\phi)$ in a given Hubble patch after $N_e$ $e$-folds of inflation. The first term in the equation describes the effect of classical slow roll while the second describes the quantum fluctuations induced by inflation.

It is convenient to define the average value of $\phi$, 
$\langle \phi\rangle$, and its variance $\Delta\phi^2=\langle(\phi-\langle\phi\rangle)^2\rangle$, where
\be
\langle F(\phi) \rangle \equiv \int_{-\infty}^{\infty} P(\phi,N_e) F(\phi) d\phi\ .
\ee
Using the Fokker-Planck Eq.~(\ref{FP}), one gets
\bea
\frac{d\langle\phi\rangle}{dN_e} &=& -\frac{1}{3H_I^2}\left\langle\frac{\partial V}{\partial\phi}\right\rangle\ ,\\
\frac{d \Delta\phi^2}{dN_e}&=&\frac{H_I^2}{4\pi^2}-\frac{2}{3H_I^2}\left\langle(\phi-\langle\phi\rangle)\frac{\partial V}{\partial \phi}
\right\rangle\ .
\eea
The first equation shows that $\langle\phi\rangle$ slow-rolls
classically down an averaged potential slope. The second equation describes the quantum spread of the field. Besides the positive contribution that makes the variance grow with $H_I$, as $\Delta\phi\sim \sqrt{N_e} H_I$, there is an effect from the potential that can eventually stop the quantum spreading
({\it e.g.}~in a potential well).

We can use these equations to gain some qualitative understanding of how the field $\phi$ will spread during its cosmological evolution (described in Section~\ref{sec:model}). During stage I, $\phi$ is stuck in a deep minimum where we can neglect quantum spreading effects. In stage II, $\phi$ rolls at the bottom of the $\phi_*$ range and it is a good approximation to assume that $\langle\phi\rangle \sim (\phi_*)_{\rm min}$, the low end of the $\phi_*$ region. The potential around that point is asymmetric, with a nearly constant and very small slope in the $\phi_*$ range and a steeper slope outside that range (the hilly side that actually slows down the field so that it tracks $(\phi_*)_{\rm min}$). As a result, the spread of $\phi$ is also asymmetric, being suppressed in the hilly side and larger inside the $\phi_*$ range. The latter spread starts growing at the beginning of stage II and, depending on how large $H_I$ is and how long stage II lasts, it might reach a steady state (with $d\Delta\phi^2/dN_e\simeq 0$). Irrespective of that, it is guaranteed
that the final quantum spread will be at most of the same order of the size of the $\phi_*$ region ($\Delta\phi_*=f/(c_\phi \epsilon)\sim f/\epsilon$) because the large barriers beyond that range suppress the spreading.\footnote{There is always a small probability that the field leaks to neighbouring minima (with the overall slope biasing the leakage towards lower minima) but the exponential suppression ($\sim\exp[-8\pi^2 V/(3H_I^4)]$) of the process overcomes the large number of e-folds in stage IV. Notice also that the minimum at which $\phi$ ends up is extremely stable against decay by quantum tunneling after inflation is over. The tunneling bounce action can be estimated (in the thin-wall approximation) to be $S_4\sim\epsilon^2f/(\Lambda g^3)\simgt (\Lambda/v)^8$, which leads to a vacuum lifetime many orders of magnitude larger than the age of the universe.} 
How far away from the 
$\phi_*$-region the barriers will stop the spreading can be estimated (by comparing the classical falling speed with the uphill quantum jumps) as 
\be
\delta \phi\sim \frac{f}{\epsilon }\times \frac{H_I^3}{g\Lambda^3}\ ,
\ee 
which is the width of the band times a factor smaller than 1, see \eq{eq:classicalrolling}. 
The same applies to stage III, with a similar bound on the final spread
$\Delta\phi\simlt  f/\epsilon$. In terms of the Higgs field expectation value, using \eq{h2phi}, this translates into an spread of the order $g \, f \Lambda/(2\lambda \epsilon)$, which is of the order of $v^2$ as we saw in the previous Section, \eq{eq:endtracking}.
So the  mechanism to solve the hierarchy problem is not spoiled.
As in \cite{Graham:2015cka}, our modified CHAIN  mechanism does not offer a solution to the cosmological constant problem.

Quantum spreading will also affect the field $\sigma$, which rolls down a much simpler potential. Its spread will grow as $\Delta \sigma \sim \sqrt{N_e} H_I$ and its final value at the end of inflation will depend on the shape of the potential near the final $\sigma$ minimum.
Concerning the evolution of the field $\phi$, the spreading of $\sigma$ will not matter (as long as classical rolling dominates): $\phi$ will simply track $\sigma$ wherever it is in a given Hubble patch.

\section{Cosmological signatures}
\label{sec:cosmo}

As we saw in the previous Sections, in contrast with the  models of Ref. \cite{Graham:2015cka}, the new-physics scale
of our model can be as large as $\Lambda\sim 10^9$\,GeV, and therefore we do not expect any new state around the weak scale.
Only  the two  additional  scalars $\sigma$  and  $\phi$ are lighter than, or at most around, the weak scale.
However, as we will see below, they are very weakly-coupled to the SM states in most of the parameter space,
and thus can only have some
phenomenological impact through astrophysical and cosmological effects.

\subsection{Properties of ${\bma \phi}$ and ${\bma\sigma}$}

We start by deriving the properties of the $\phi$ and $\sigma$ scalars.
After the slow-rolling process ends and $\sigma$ settles in a minimum, no cancellation is expected
in the $A(\phi,\sigma,H)$ amplitude, so that $A(\phi,\sigma,H) \sim \epsilon \Lambda^4$.
The mass of $\phi$ is thus controlled by $A \cos(\phi/f)$  and can be estimated as
\be
m^2_{\phi} \sim \frac{\epsilon \Lambda^4}{f^2} \sim  g \frac{\Lambda^5}{f v^2}  \lesssim v^2\, ,
\ee
where we used \eq{eq:quantumstability} and \eq{eq:endtracking}  to obtain the second equality and the
upper bound on $m_\phi$.
For $\pp$ we expect that higher-order terms in $g_\pp \pp/\Lambda$, not shown for simplicity in \eq{eq:V}, 
 give it a mass of order
\be
m^2_{\sigma} \sim  g_\sigma^2 \Lambda^2 \ll m^2_{\phi}\, .
\ee
In the allowed part of the parameter space of our model the masses of the two scalars can change by many orders of magnitude,
spanning the range $[10^{-20}, 100]$\,GeV for $\phi$ and $[10^{-45}, 10^{-2}]$\,GeV for $\sigma$.
Contours of constant $m_{\phi}$ and $m_{\sigma}$ are shown in Fig.~\ref{fig:gversusLambda}. 

These two scalars  interact with the SM particles mainly through a mass mixing with the Higgs.
The corresponding mixing angles can be estimated as
\be
\theta_{\phi h}  \sim  \frac{g \Lambda v }{m_h^2}\ ,\ \ \
\theta_{\sigma \phi }  \sim  \frac{g_{\sigma} f v^2 }{\Lambda^3}\ , \  \ \
\theta_{\sigma h}  \sim {\rm Max}\left\{ \theta_{\sigma \phi } \theta_{\phi h}\ ,\
\frac{g^2}{16\pi^2}\frac{g_\pp \Lambda^7}{f^2 v^3m^2_h} \right\}\, .
  \label{eq:hsigmamixing}
\ee
Notice that the $\phi-h$ mass mixing coming from  $\partial^2_{ \phi h}V\sim \epsilon \Lambda^2 (v/f) \sin(\phi/f)$
is suppressed  at the minimum where  we have  $\sin (\langle \phi\rangle /f)\sim gf/(\epsilon \Lambda)\sim v^2/\Lambda^2\ll 1$.\footnote{This is to be contrasted with the beginning of Phase IV 
when $\sin (\phi/f) \sim 1$, as used to derive \eq{eq:endtracking},
since barriers are smaller at this earlier stage.
At the end of Phase IV the barriers have grown large, and $\phi$ is 
close to the minimum of its cosine potential.}
The first contribution in $\theta_{\sigma h}$  arises at tree-level,
whereas the second one originates from a $\phi$-loop. 
For most of the parameter space we consider, this loop term dominates over the tree level one.
The scalar potential  \eq{eq:V}  also gives rise to interactions between $\phi$ and the Higgs, not suppressed by the
small mixing angle $\theta_{\phi h}$, that are of order
 \be
  \label{eq:phihcouplings}
 \phi\phi hh : \  \   \epsilon \Lambda^2/f^2  \ ,  \ \ \ \ \
  \phi \phi h  : \   \  \epsilon v \Lambda^2/f^2\, ,
 \ee
and will play an important role in the thermal production of $\phi$.
The decays of  $\phi$ and $\sigma$  are mediated by the mixing with the Higgs, and thus the
widths are given by
\bea
\Gamma_{\phi} \sim \theta^2_{\phi h} \Gamma_h(m_{\phi})\ ,\quad
\Gamma_{\sigma} \sim \theta_{\sigma h}^2 \Gamma_h (m_{\sigma})\, ,
\eea
where $\Gamma_h (m_i)$ is the SM Higgs width evaluated at $m_h=m_i$.
Contours for  $\Gamma_{\phi, \pp}$ are shown in Fig.~\ref{fig:gversusLambda} (the values of the width $\Gamma_h(m_i)$ are subject to large theoretical uncertainties in the mass region around 1\,GeV where several hadronic decay channels open up~\cite{Ellis:1975ap}; we used the expressions given in Ref.~\cite{Bezrukov:2009yw} --see also Refs.~\cite{Schmidt-Hoberg:2013hba,Clarke:2013aya}).
For masses  below $2m_e\sim 1$~MeV,  we have
$ \Gamma_h(m_{i}) \sim 
\left({m_{i}}/{m_h}\right)^3 \Gamma_{h\rightarrow\gamma \gamma} (m_h)  $, and
therefore, in a major part of  the parameter space, $\phi$ and $\sigma$ have suppressed decay widths controlled by the decay into photon pairs. 
As shown in Fig.~\ref{fig:gversusLambda}, there is a sizable part of the parameter space in which $\phi$ is
cosmologically unstable ($\Gamma_\phi > H_0$, where $H_0$  is the present Hubble value), but sufficiently long-lived to decay after Big Bang Nucleosynthesis (BBN)
($\Gamma_\phi < H_{BBN}\equiv H(T=1  \mbox{ MeV})$). As we will see in the following, this region of the parameter space can be constrained by cosmology. On the other hand, $\sigma$ is cosmologically stable in most of the relevant parameter space,
and can decay within the age of the universe only in a small corner of the parameter space, namely
for $g_\sigma \gtrsim 10^{-8}$ and $\Lambda \lesssim 10^4$\,GeV.

We can now easily estimate the cosmological abundances of  $\phi$ and $\sigma$, either   stored in late classical oscillations (vacuum misalignment) or  from thermal production. This will allow us to set  bounds on the model from overclosure of the universe, post-BBN decays or astrophysical 
constraints.

\subsection{Impact of ${\bma\phi}$ and ${\bma\sigma}$ on standard cosmological predictions}
\label{sec:cosmo}
 
In this work we assume for simplicity that, once both $\phi$ and $\sigma$ have settled in their minima, inflation ends with an unspecified reheating period. We will assume a reheating temperature
higher than the EW scale in what follows.



 \subsubsection*{Abundances of ${\bma\phi}$ and ${\bma\sigma}$ from vacuum misalignment}
 
If after inflation and reheating,  the fields $\phi$ and $\sigma$
 end up displaced from their minima, they     will 
  fall towards them,  oscillating around them if their lifetimes are large. The energy density stored in the field oscillations behaves like cold dark matter and can potentially overclose the universe today or dissociate light elements if the decay takes place  during or after BBN. 
At the start, the field energy density  is dominated by the potential energy, but as the fields roll to their minima and pick up speed, the kinetic energy grows. When both contributions are of similar size, 
 one can properly talk about oscillating fields.
This  happens for the  Hubble value  $H\lesssim m_i$ ($i=\phi,\pp$)  or, equivalently, in a radiation-dominated universe, at $T^{i}_{osc}\sim \sqrt{m_{i} M_P}$.
 In that oscillating regime, the energy density scales as that of non-relativistic matter, $\rho_i (T)\sim \rho^{i}_{ini} (T/T^{i}_{osc})^3$,
where $\rho^i_{ini}$ is the initial amount of energy at the start of the oscillating regime (this is smaller than the original potential energy density, but only by an order one factor).

Let us start considering  the $\sigma$ field.
We expect that during inflation $\sigma$ slowly rolled down to its global minimum, somewhere in its $\sim \Lambda/g_{\sigma}$ range, as  this requires a number of e-folds similar to the $N_e$ estimated in Eq.~(\ref{eq:longinflation}).
Because of quantum effects (see Section~\ref{sec:spread}), $\sigma$ reached the minimum with a spread $\sqrt N_e H_I$.
We can use this result to estimate the typical displacement from the minimum at the end of inflation:
$$
(\Delta \sigma)_{ini} \sim \sqrt N_e H_I\ .
$$
This displacement is quite large though still smaller than $\Lambda/g_{\sigma}$, and hence we can estimate the amount of
energy density as $\rho^{\sigma}_{ini} \sim m_{\sigma}^2 (\Delta \sigma)_{ini}^2$.
Using $N_e\sim H_I^2/(\Lambda^2 g_\pp^2)$, we finally get
$\rho^{\sigma}_{ini} \sim H_I^4$.
The energy density stored in $\sigma$ oscillations today, relative to the critical energy density, is then, using \eq{eq:inflation},
$\Omega_{\sigma} \simgt 
\left({4 \times10^{-27}}/{g_\sigma}\right)^{3/2}\left({\Lambda}/{10^8\,\mbox{GeV}}\right)^{13/2}\,.$
%
The bound to avoid universe overclosure 
($\Omega_{\sigma} \lesssim 1$), translates into 
\be
g_{\sigma} \gtrsim 4\times 10^{-27} \left( \frac{\Lambda}{10^8\,\mbox{GeV}} \right)^{13/3}\,.
\label{eq:overclosure}
\ee
The contour $\Omega_{\sigma} = \Omega_{DM}$ is plotted in Fig.~\ref{fig:gversusLambda}, limiting the excluded blue region.
It shows that $\sigma$ can be a good dark matter candidate in the region where
the bound~(\ref{eq:overclosure}) is saturated, in particular at large $\Lambda$. 
For certain values of $m_\pp$, there can be other cosmological constraints.
For example,   for $\Omega_{\sigma }\gtrsim \Omega_{DM}/20$, the mass range  $10^{-32} \mbox{ eV }\lesssim  m_{\sigma} \lesssim 10^{-25.5}$ eV is excluded by structure formation \cite{Hlozek:2014lca},
 while masses around  $m_{\sigma} \sim 10^{-11}$ eV may be constrained by Black Hole superradiance \cite{Arvanitaki:2014wva}.
Interestingly,  
for  the particular case $m_{\sigma} \sim 10^{-24}$ eV, $\sigma$ could be searched for by the SKA pulsar timing array experiment \cite{Khmelnitsky:2013lxt}.
Let us finally notice that  there are   ways to go around the bound~(\ref{eq:overclosure}),  for instance,  by assuming   a late entropy production after $\sigma$ has started to oscillate, as can occur  if reheating is a very slow process such that  $T_{RH}<T^\pp_{osc}$ \cite{Giudice:2000ex}.

The situation for $\phi$ is rather different. At the end of its evolution,
$\phi$ is trapped  in a region with high barriers
and its displacement from the minimum originates  from the possible quantum spreading. 
The initial energy density arising  from this  displacement  was at most
$\rho^{\phi}_{ini} \sim H_I^4$, that, since $m_\phi\gg m_\pp$ and then $T^\phi_{osc}\gg T^\pp_{osc}$, 
gives today a completely negligible effect.

\subsubsection*{Thermal production of ${\bma\phi}$}

Thermal production of $\phi$  arises mainly from  the couplings of 
\eq{eq:phihcouplings}.
In particular, from the $\phi\phi hh$-coupling
we can have   double-production
from the thermal bath  via
$ h + h \rightarrow \phi + \phi $.
\footnote{Double production can also be mediated by the process
$SM + SM \to h^{(*)}\to \phi+\phi$  induced by
the $\phi\phi h$-coupling, which can lead to a similar thermal production as the one discussed here.
Single production,  on the other hand,
is due to interactions that are linear in the $\phi$ field and are thus suppressed by the small mixing angle $\theta_{\phi h}$,
 and can be neglected.} 
At   $T\gtrsim m_h$,  this double-production cross-section is estimated to be $ \langle \sigma_A {\rm v}\rangle \sim \epsilon^2 (\Lambda^4/f^4)/T^2$. This implies that $\phi$ can reach thermal equilibrium only for $T$ in the interval $[m_h,\epsilon^2 M_P (\Lambda/f)^4]$, in which the $\phi$ production rate is faster
than the rate of expansion. 
This region corresponds roughly to the area above  the $\Gamma_\phi=H_{BBN}$ line of Fig.~\ref{fig:gversusLambda}, 
so we conclude that in most of the parameter space, $\phi$ 
never thermalizes.\footnote{This also implies that  we can neglect thermal corrections to the potential for $\phi$ in the analysis of its cosmological evolution.}

The number density of $\phi$ produced thermally is obtained by solving the Boltzmann equation 
\be
\frac{dn_{\phi}}{dt} + 3 H n_{\phi} = -\langle \sigma_A {\rm v} \rangle (n^2_{\phi}-{n^2_{\phi,eq}})\,,
\label{eq:Boltzmann}
\ee
where $n_{\phi,eq}$  is the equilibrium number-density of $\phi$. 
This equation can be conveniently re-written in terms of the dimensionless quantities $x=m_{\phi}/T$ and $Y_\phi=n_{\phi}/s$, where $s$ is the entropy per comoving volume, $s= 2\pi^2g_{*s}T^3/45$. Assuming a radiation-dominated era, with energy density $\rho_R= \pi^2 g_* T^4/30$  (here, $g_{*}\sim g_{*s}\sim 100$ counts the  number of relativistic degrees of freedom)
and using that $Y_\phi \ll Y_{\phi, eq}$ in the large portion of parameter space in which $\phi$ does not thermalize, one gets: 
\be\label{eq:dY}
\frac{dY_{\phi}}{dx} \simeq \frac{\langle \sigma_A {\rm v}\rangle  {\cal C} m_{\phi} M_P}{x^2}Y^2_{\phi,eq}\,    ,
\ee
where ${\cal C}= 2 \pi \sqrt{90}g_{*s}/(45\sqrt{g_*})\simeq 13.7$.  
For relativistic $\phi$, $x\ll1$, the equilibrium density is approximately given by $Y_{eq}\sim 0.278/g_{*s}$. 
This leads to the approximate formula
\be
Y_\phi(T)\sim \epsilon^2 \frac{\Lambda^4}{f^4} {\cal C}Y_{\phi, eq}^2 \frac{M_P}{T}\,.
\label{eq:phiabundance}
\ee
The $\phi$ production is maximal at $T\sim m_h$.
Using \eq{eq:phiabundance} evaluated at $T\sim m_h$  and assuming no late entropy production, we 
can deduce, in the parameter region where $\phi$ is cosmologically stable, the contribution of $\phi$ to dark matter today,  $\Omega_{\phi} \sim m_{\phi} Y_{\phi} {s_0}/{\rho_c} $ where $s_0$ is the present entropy density.
We find that $\Gamma_{\phi}$ is subdominant; for example,  along the contour $\Gamma_{\phi}=H_0$ 
in Fig.~\ref{fig:gversusLambda}, we  have $ \Omega_{\phi}  \lesssim 3 \times 10^{-4}$, while  for   $\Gamma_{\phi}\simeq 10^{-10}H_0$ we find  $ \Omega_{\phi}  \lesssim  7 \times 10^{-11}$.

\subsubsection*{Constraints from BBN and Gamma-Ray observations}

As mentioned earlier, there is a region of parameter space in which $\phi$ is not
cosmologically stable and decays after BBN.
This is problematic if the decay of $\phi$ injects into the thermal bath an energy per baryon $E_{p.b} \gtrsim O$(MeV), leading to a modification of the predictions for the abundances of the light elements. Since 
$E_{p.b}\sim m_{\phi} Y_{\phi} n_{\gamma}/n_b$, where $Y_{\phi}$ is the abundance prior to decay (i.e. the value of \eq{eq:phiabundance}  obtained as if $\phi$ were stable), this  results in the bound
 $m_{\phi}Y_{\phi}\lesssim 10^{-12}$\,GeV, which however could be weakened sensitively depending on the precise value of the lifetime \cite{Cyburt:2002uv}. 
In the parameter space region where $\phi$ decays between cosmic times $\sim 1$ s and $ H_0^{-1}\sim 10^{17}$ s, $m_{\phi}$ varies from $\sim$ 100\,GeV down to 1 MeV and $m_{\phi} Y_{\phi}$ varies from $10^{-2}$ to $10^{-12}$\,GeV.  Above the di-pion threshold, hadronic decays constrain short lifetimes $\Gamma^{-1}_{\phi} \lesssim 10^3$ s while for lighter $\phi$, electromagnetic decays  constrain large lifetimes $\gtrsim 10^5$ s \cite{An:2014twa,Kawasaki:2004qu}.
In addition, the Cosmic Microwave Background (CMB) constrains lifetimes $\sim [10^{10} - 10^{13}]$ s for $E_{p.b}$ down to $O$(eV). 
Therefore,  we expect that most   of the region of the parameter  space  delimited by the lines $\Gamma_{\phi}=H_{BBN}$ and  $\Gamma_{\phi}=H_0$ in Fig.~\ref{fig:gversusLambda}  is excluded. A dedicated analysis is however   needed to derive the precise excluded regions that is beyond the scope of this paper.

On the other hand, for regions in which the $\phi$ lifetime is larger than the age of the universe,
there are    strong constraints  coming from  decays generating a  distortion 
in the galactic and extra-galactic diffuse X-ray or gamma-ray background.
 In particular, sub-GeV dark matter decaying into photons 
should satisfy  $\tau_{DM} \gtrsim10^{27}$ s  \cite{Essig:2013goa}.
Since the gamma-ray flux scales as 
${d\Phi_{\gamma}}/{dE} \propto Y_{\phi} \Gamma_{\phi}$, we can translate this bound  into
$\tau_{\phi} > 10^{27} \mbox{ s }\times {\Omega_{\phi}}/{\Omega_{DM}}$.
 This  excludes   the thin brown band   of Fig.~\ref{fig:gversusLambda}.
 
We stress again that  the cosmological constraints derived above  can be evaded  if the
 temperature of the universe never reaches  $m_h$, in which case 
 the thermal production of $\phi$ is  suppressed.


\section{Conclusion}
\label{sec:conclusion}

The aim of this paper has been to provide an existence proof of a model,
based on the idea of  \cite{Graham:2015cka}, 
that can naturally accommodate a small electroweak scale without (sadly, to say) requiring visible new-physics
at present and  far future colliders.
The model is based on a cosmological evolution of the Higgs and two axion-like states 
whose back-reactions lead  to a naturally small electroweak scale.
The only new-physics   of the model    consists of these  two   scalars, $\phi$ and $\sigma$,
that in most of the parameter space are very light and very weakly coupled to the SM.
Therefore strategies to detect them are completely different, as they do not require 
powerful high-energy colliders, but dedicated searches in the sub-GeV regime.

Interestingly, one of the two light states, namely $\pp$, could be a dark matter candidate.
Because of its small mass, $\sigma$ starts oscillating very late and can therefore make a large contribution to dark matter today.  The field $\phi$ cannot contribute to more than $\Omega_{\phi} \lesssim 10^{-10}$. For this maximum value, it may still be detected in gamma-ray observations from its late decay.

Part of the parameter space of our model can be tested through observations of the diffuse gamma-ray backgrounds, black hole superradiance and even in pulsar timing arrays. In addition, there is a rather rich BBN and CMB phenomenology which motivates a more thorough study. Unfortunately,  fifth-force signals and Equivalence Principle violations in intermediate mass ranges seem too small to be seen in the near future.

 There are important discussions that we leave for the future.
An interesting aspect of our   mechanism is that it does not depend on initial conditions for the new scalars, provided that it takes place during a long period of inflation.
Similarly to   \cite{Graham:2015cka}, we need a very long period of inflation,  $N_e \gtrsim 10^{25}$, and an inflation scale rather low,  $\lesssim 10^9$ GeV. In addition, inflation requires super-Planckian field excursions. All of these issues deserve further study.  It seems clear that the possibility to further increase the effective cutoff $\Lambda $ might help with these questions.

To close on a more philosophical note, the ideas proposed in \cite{Graham:2015cka} and pursued here represent a new twist in the long and  fruitful history of the interplay between particle physics and cosmology. While in the past particle physics has been a crucial ingredient to understand the cosmological history of our universe, if these new ideas were correct, cosmological evolution would be a key ingredient in the understanding of some key parameters of particle physics.

\section*{Acknowledgments}
We thank Joan Elias-Mir\'o, Eduard Mass\'o and Marc Riembau  for very useful discussions in the early stages of this project and David~E.~Kaplan, Kai~Schmidt-Hober  and Alexander~Westphal for very interesting discussions. 
We acknowledge support by the Spanish Ministry MEC under grants FPA2014-55613-P, FPA2013-44773-P, and FPA2011-25948, by the Generalitat de Catalunya grant 2014-SGR-1450 and by the Severo Ochoa excellence program of MINECO (grant SO-2012-0234).
CG is supported by the European Commission through the Marie Curie Career Integration Grant 631962.
CG and GS are supported by the Helmholtz Association. The work of AP has  been supported by the Catalan ICREA Academia Program.

\appendix
\section{A partial UV completion of  the model}

Let us start the discussion by summarizing the basic building blocks of the potential of \eq{eq:V}.
The first ingredient is the effective potential for the scalar $\phi$ that contains two types of terms.
The periodic term,\footnote{The form of the periodic function is not important for the model.
The choice $\cos(\phi/f)$ has been selected in view of a possible axion-like UV completion, as we will see in the following.}
associated to a discrete shift invariance,
$\phi \rightarrow \phi + 2 \pi f$,
and the linear terms  that break this symmetry by the small coupling $g$.
It is crucial that both types of terms couple to $|H|^2$ for the CHAIN  mechanism to work.
Due to  \eq{eq:endtracking}, we have $g\ll \epsilon$ 
and therefore the   linear   terms are  always smaller than the  periodic term.
In addition, the model also contains a second field, $\sigma$, whose  
interactions are controlled by  the small coupling $g_\sigma$, in such a way that $\sigma$ always appears in the
combination $g_\sigma \sigma/\Lambda$. In analogy to the $\phi$ field, we can
associate $\sigma$ to a continuous shift invariance, whose breaking is controlled by $g_\sigma$.
It is crucial to have a coupling of $\pp$ to the periodic function of $\phi$,
but not a direct coupling  to the Higgs that would spoil the CHAIN mechanism, as we mentioned before.
Quantum effects via a loop of $\phi$-field can of course generate this coupling, $\sigma |H|^2$, but its size of order $O(\epsilon^2 g_\sigma)$ is small enough and does not destabilize the weak scale.

\begin{figure}
\centering
\includegraphics[width=.28\textwidth]{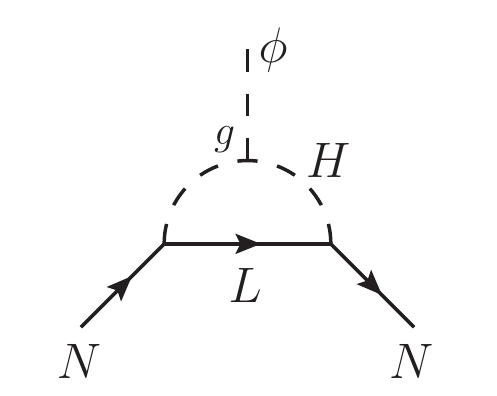}
\includegraphics[width=.28\textwidth]{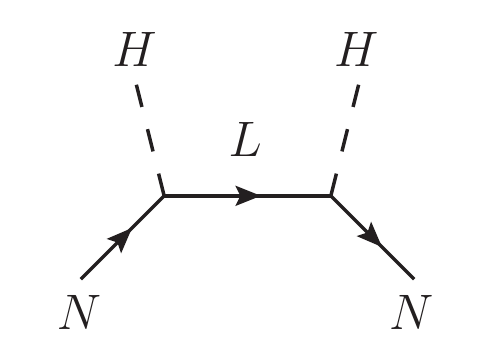} 
\includegraphics[width=.38\textwidth]{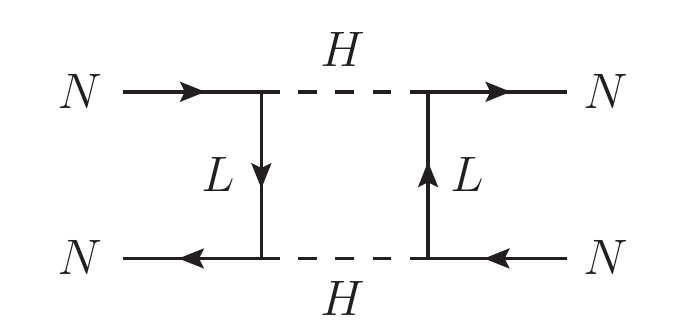}
\caption{\emph{{\bf Left:} Diagram generating $\phi \overline NN$  at the radiative level.
{\bf Middle:} Diagram contributing to the coupling $\overline NN|H|^2$.
 {\bf Right:}
Diagram generating an $O(\epsilon^2)$ contribution to  $(\overline NN)^2$.}}
\label{fig:loop}
\end{figure}

Although each of these terms needs an appropriate UV origin,
here  we are only interested 
in   providing a UV model that explains the generation of the
periodic terms in the  potential of \eq{eq:V}.
The model is a generalization of the
non-QCD model proposed in Section~\ref{sec:Consistency} of Ref.~\cite{Graham:2015cka}.
It consists of a   gauge sector, which, analogously to QCD,  we  take to be $SU(N)$, with 
  two Dirac fermions, $L$ and $N$, belonging to the fundamental representation of $SU(N)$.
Under the $SU(2)_L \times U(1)_Y$ SM group,  $L$ has the quantum numbers of
a lepton doublet, while $N$ is a singlet.
We assume  that the $SU(N)$ gauge sector  becomes   strongly-coupled at the scale $\Lambda$.
A key ingredient of the  model is the presence of a specific set of mass and interaction terms for the fermions
that break the accidental  global symmetries.
We assume that  the $L$ and $N$ fields have Dirac masses  (here and in the following we neglect $O(1)$ parameters):
\begin{equation}
{\mathcal L}_{mass} = \Lambda \overline L L + \epsilon \Lambda \overline N N\, ,
\end{equation}
and  couplings to the SM Higgs given  by
\begin{equation}
{\mathcal L}_{Yuk} = 
\sqrt{\epsilon}\, \overline L H N+h.c.\,.
\end{equation}
Finally, interaction terms of the singlet $N$ to the $\sigma$ and $\phi$ fields are included with couplings
of order $\epsilon g$ and $\epsilon g_\sigma$ respectively
\begin{equation}
{\mathcal L}_{N} = \epsilon g \phi \overline N N + \epsilon g_\sigma \sigma \overline N N\, .
\end{equation}
As can be seen from the Lagrangian above, 
we have associated to each $N$ field 
a coupling $\sqrt{\epsilon}\ll 1$.
In the limit $\epsilon \rightarrow 0$  the theory acquires an additional chiral invariance (broken only by
the  axial anomaly).
It is interesting to notice that even if  we  do not introduce in the Lagrangian the coupling of the $\phi$ field to $N$, it
is nevertheless generated at the radiative level due to the presence of the $g \Lambda \phi |H|^2$ coupling in the
effective Lagrangian, as shown by the left diagram of   Fig.~\ref{fig:loop}.

We also assume  
that the  $\phi$ field interactions are invariant under a shift-symmetry,  $\phi\to \phi+c$, up to the explicit breakings
due to $g$, and  an  anomalous interaction term 
\be
\frac{\phi}{f}  G'_{\mu\nu} \widetilde G'^{\mu\nu}\, ,
\label{anom}
\ee
 where $G'$ denotes the $SU(N)$  field strength.
Analogously to the QCD axion, the field $\phi$ acquires a periodic effective potential as a consequence of the
chiral anomaly.   
The best way to estimate  the size of this contribution is to perform a chiral rotation for $N$ such that
one eliminates the term~(\ref{anom}) but  generates $m_N\overline N N\to m_N e^{i\phi/f}\overline N N$,
where $m_N$ is the effective mass of $N$:
\begin{equation}
m_{N} \simeq \epsilon \left(\Lambda + g_\sigma \sigma + g \phi -\frac{|H|^2}{\Lambda}\right)\, ,
\label{uvmass}
\end{equation}
with the last term arising from the middle diagram of   Fig.~\ref{fig:loop}.
Using $\langle\overline N N\rangle \sim \Lambda^3$, the term $m_N e^{i\phi/f}\overline N N+h.c.$ gives,  at $O(\epsilon)$,
\begin{equation}
V \simeq \Lambda^3 m_N \cos (\phi/f)\, .
\label{uvpot}
\end{equation}
Equation~(\ref{uvpot}), together with \eq{uvmass}, reproduces the periodic term of \eq{eq:V}.

Using this explicit UV model we can also analyze possible additional contributions to the $\phi$ potential.
For example, at $O(\epsilon^2)$
we expect contributions to the $\phi$ potential coming from terms $(\overline N N)^2$ generated at the quantum level from
diagrams as the one   shown in Fig.~\ref{fig:loop}.
After the chiral rotation described above, that eliminates \eq{anom}, we  have
$(\overline N N)^2\to (\overline N Ne^{i\phi/f})^2$, which leads to
a term for the $\phi$ potential $\sim \epsilon^2 \Lambda^4 \cos^2 (\phi/f)$.
As we discussed in the main text, in order
for the relaxation mechanism to work we need to suppress these terms 
with respect to the leading potential in \eq{eq:V}. This leads  to   the condition  in Eq.~(\ref{eq:quantumstability}).

\section{Relaxation in a two Higgs doublet scenario}\label{app:TwoHiggsDoublet}

In this appendix we present a relaxation model based on a two Higgs doublet model (2HDM).
The motivation for this is,
 as mentioned in the
introduction,   to generate the term $h\cos(\phi/f)$
that requires a  second source of EWSB.
If the second Higgs is also elementary,  we must find a way to 
keep its mass  also small.   Otherwise, 
we  expect that,
at the end of the relaxation process, only one Higgs being light, while the second doublet 
having   generically  a mass of the order $\Lambda$.

To solve this problem we can advocate an additional global $SU(2)_R$ invariance at the scale $\Lambda$ under which the two Higgses transform as a doublet,  $(H_1,H_2)$, ensuring that both  have the
same masses and quartic couplings. This symmetry guarantees that the masses of the two Higgses are canceled
simultaneously by the $\phi$ field.
The $SU(2)_R$ symmetry can be easily extended to the third quark generation sector by considering a type-II 2HDM
 in which the $b_R$ and $t_R$ components form an $SU(2)_R$-doublet, and the Yukawa term is given by
\be
  y\, \bar Q_L  (H_1\   H_2)\, (b_R\  t_R)^T+h.c.\, .
  \label{yuka}
\ee
Below $\Lambda$,  the most important  source of breaking of the $SU(2)_R$ invariance 
arises from the SM hypercharge gauge-boson that of course distinguishes  $b_R$ from $t_R$ (and $H_1$ from $H_2$).
\footnote{An additional breaking is induced at one loop by light fermion Yukawas. These effects are however subleading with respect to the ones induced by the hypercharge gauging.}
The effective potential involving the Higgses and the  field $\phi$ reads
\begin{eqnarray}
V(H_1, H_2, \phi) &=& \Lambda^3 g \phi +m^2_H(\phi)  \left(|H_1|^2 + |H_2|^2\right) + \lambda \left(|H_1|^2 + |H_2|^2\right)^2\nonumber\\
&&+\; \Delta m^2  |H_1|^2 + (\epsilon \Lambda^2 H_1 H_2 e^{i\phi/f}+h.c)+\cdots\, ,\label{eq:pot2HD}
\end{eqnarray}
where $m^2_H(\phi)=-\Lambda^2 \left(1 - g \phi/\Lambda\right)$.
In the above expression we included the leading $SU(2)_R$-breaking  $\Delta m^2 |H_1|^2$, 
that arises from   two-loop diagrams  involving  fermions and  the hypercharge gauge-boson,
and whose size can be estimated to be 
\begin{equation}
\Delta m^2 \sim \frac{g'^2 y^2}{(16 \pi^2)^2} \Lambda^2\,.
\label{twoloop}
\end{equation}
If we want this model to have two independent sources of EWSB,   $H_1$ and $H_2$, 
we encounter  a phenomenological problem.
The model will contain  extra light Higgs, of mass of order $\lesssim \sqrt{\lambda} v$, that have not been seen at the LHC.
For this reason, we are forced to  look for  2HDMs in which only one Higgs, $H_2$, is responsible for
triggering EWSB, while the other Higgs, $H_1$,  gets a vacuum expectation value  from its mixing with $H_2$.
The masses of the extra Higgs will then be of order $\Delta m$ that can be heavier than the EW scale
and escape present detection.

The dynamics of this model is quite straightforward to understand. When, as a consequence of the slow-roll of $\phi$,
$m^2_H(\phi)$ becomes negative, only $H_2$ turns on,
as the $H_1$ mass is still positive if, as we assume from now on, $m^2_H(\phi)+\Delta m^2>0$.
Nevertheless, the periodic term in the potential  automatically induces a tadpole for $H_1$
 which displaces it from the origin. 
In such a scenario  we have
\begin{equation}
\langle H_2 \rangle \simeq v\,, \qquad \langle H_1 \rangle \simeq  \frac{\epsilon\Lambda^2}{\Delta m^2} v\,.\label{eq:VEVs}
\end{equation}
As the 3rd family Yukawas are equal, \eq{yuka}, 
we have  $m_b/m_t\simeq \langle H_1 \rangle/\langle H_2 \rangle$,  that requires
\be
\epsilon\Lambda^2/\Delta m^2 \simeq m_b/m_t\, .
\label{ratiom}
\ee
In order for the relaxation mechanism to work, we need to ensure that the potential is radiatively stable. The fact 
that we have two Higgses helps to protect the periodic term of \eq{eq:pot2HD}  that can only be self-renormalized.
Therefore the only  one-loop extra term  that we can have is given by 
\begin{equation}
\frac{1}{16 \pi^2} \epsilon^2 \Lambda^4 \cos^2 (\phi/f)\,,
\end{equation}
that  should be smaller than the periodic term in Eq.~(\ref{eq:pot2HD}) (recall that here, contrary to our main model, we are not introducing the field $\sigma$).
This leads to the condition
\begin{equation}
\epsilon \lesssim16 \pi^2 \frac{m_b}{m_t} \frac{v^2}{\Lambda^2}\,,
\end{equation}
which, together with  \eq{twoloop}  and \eq{ratiom}  gives an upper bound on the cut-off $\Lambda$:
\begin{equation}
\Lambda \lesssim \frac{(16 \pi^2)^{3/2}}{g' y} v \simeq 10^6\,\mathrm{GeV}\,.
\end{equation}
This model is thus characterized by three scales: the weak scale $v$,
the mass of the second Higgs doublet $m_{H_1} \sim \Delta m\lesssim 10\ v$, and the scale of new dynamics $\Lambda \lesssim 10^3\ v$. This implies that only the second Higgs doublet would be reachable
at the  next LHC runs.

\end{document}